\documentclass[twocolumn,showpacs,preprintnumbers,amsmath,amssymb,pra]{revtex4}

\usepackage{bm}%
\usepackage{graphicx}
\usepackage{dcolumn}
\usepackage{natbib}

\begin{document}

\title{Improving resolution by means of ghost imaging}

\author{Pengli Zhang, Wenlin Gong, Xia Shen, Dajie Huang and Shensheng Han}

\email{sshan@mail.shcnc.ac.cn}

\affiliation{ Key Laboratory for Quantum Optics and Center for Cold
Atom Physics, Shanghai Institute of Optics and Fine Mechanics,
 Chinese Academy of Sciences, Shanghai 201800, China
}

\date{\today}

\begin{abstract}
As one of important analysis tools, microscopes with high spatial
resolution are indispensable for scientific research and medical
diagnosis, and much attention is always focused on the improvement
of resolution. Over the past decade, a novel technique called ghost
imaging has been developed that may provide a new approach toward
increasing the resolution of an imaging system. In this paper, we
introduce this technique into microscopes for the first time and
report a proof-of-principle experimental demonstration of a
microscope scheme based on ghost imaging.
\end{abstract}


\maketitle


  During the past half century some sophisticated optical technologies, such as confocal
microscopes \cite{CLSM,Car85}, transmission x-ray microscopes
\cite{Nie76,Sch94} and so on, have been exploited to achieve
excellent resolution. For a lens-based optical microscope, the
resolution is determined by the extent of the point spread function,
and the extent primarily depends on the wavelength of illumination
light and the numerical aperture (NA) of the objective lens
\cite{Abble73}. A lens with high NA is one of the key factors of
realizing high resolution. However, some practical conditions may
restrict the use of a high-NA lens. For example, medical endoscopes
\cite{Endo} examining human internal organs require lenses with
small aperture and transmission x-ray microscopes \cite{Jaco92}
detecting thick specimens demand Fresnel zone plates with long focal
depth, which both limit NAs of the lenses. The development
\cite{pitman95,Bennink02,cheng04,Bache04,Valencia05,Wulinan05,
Cheng07,Gatti06,Liu07,Cai05,Angelo05,Zhang07} of ghost imaging in
recent ten years, now brings a new way to increase the resolution of
these lens-limited microscopes.

   Ghost imaging is a technique that forms an image of an object by
measuring two correlated optical fields with the use of entangled
sources \cite{pitman95} or ``classical" sources, such as pairs of
momentum-correlated laser pulses \cite{Bennink02} and thermal light
\cite{Valencia05,Wulinan05}. In general, a conventional imaging
system only needs one detector to record the intensity distribution
related to the amplitude and phase of a target object. In quantum
theory of photodetection, the light intensity measured by the
detector can be represented by the first order correlation function
\cite{Glauber}:
\begin{equation}\label{Eq01}
G^{(1)}(x,t)=<E^{(-)}(x,t)E^{(+)}(x,t)>,
\end{equation}
where $E^{(\pm)}(x,t)$ are the quantized positive and negative
frequency parts of the field at space-time location (x,t). While a
ghost imaging system must simultaneously record the intensities of
two correlated beams: a beam that travels a path (the test arm)
including the object and the other beam that passes through a
reference optical system (the reference arm), the information about
the object is exacted from the correlation between two recorded
intensities. The correlation can be evaluated through the second
order correlation function \cite{Glauber}:
\begin{eqnarray}\label{Eq02}
&G&^{(2)}(x_{1},x_{2},t,t)=\nonumber\\
&<&E^{(-)}(x_{1},t)E^{(-)}(x_{2},t)E^{(+)}(x_{1},t)E^{(+)}(x_{2},t)>
\end{eqnarray}
where $E^{(\pm)}(x_{1},t)$ and $E^{(\pm)}(x_{2},t)$ are the field
operators in two detecting planes at the same time. The unique work
principle of ghost imaging leads to some interesting optical
phenomena, such as reconstructing a ``ghost" image in the reference
arm while the measured object is in the test arm
\cite{pitman95,Bennink02,Valencia05,Wulinan05}, implementing
coherent and incoherent imaging in the same system only by changing
the detection modes \cite{Bache04}, and lensless Fourier-transform
imaging with thermal light \cite{cheng04}, that the conventional
imaging system can't realize.
\begin{figure}[htb]\
 \centering\includegraphics[width=7cm]{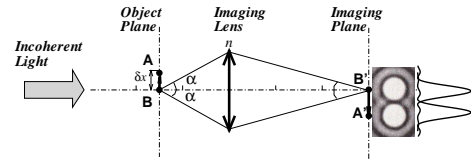}
\caption{Schematic of a simple conventional imaging system.
\textit{n} is refractive index of the medium, and \textit{$\alpha$}
is the half angle of the cone of light acceptable by the imaging
lens.} \label{fig1}
\end{figure}

  A schematic of a simple conventional imaging system is shown in Fig.\ref{fig1}.
Under incoherent illumination, the image of a point at the object is
not infinitely small, but is a circular diffraction image, or called
diffraction spot. The width of the spot represents the resolution of
the image. According to the Rayleigh criterion \cite{Rayleigh}, the
resolution limit of the system in the object plane is determined by
\begin{equation}\label{Eq1}
\delta x=0.61\frac{\lambda}{n\sin(\alpha)}
\end{equation}
\noindent where $n\sin(\alpha)$ denotes the NA of the imaging lens
and $\lambda$ is the wavelength of light. Eq.(\ref{Eq1}) shows that
the improvement of resolution relies on shorter wavelength and
higher NA. While the wavelength is given and the lens is limited,
it's still desirable to obtain high-resolution images. To achieve
this goal, we apply ghost imaging technique into the conventional
imaging system and present the theoretical and experimental
demonstration of a new microscope scheme.

\begin{figure}[htb]
\centering\includegraphics[width=7cm]{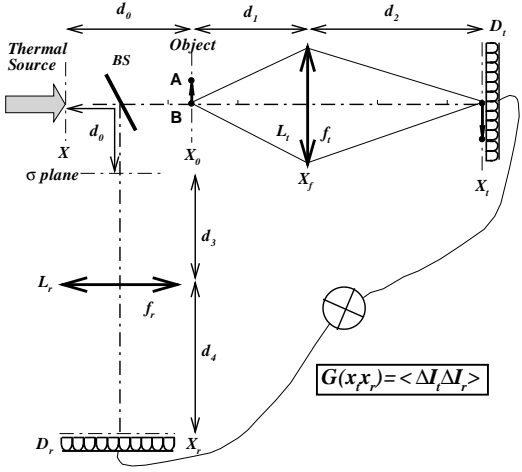} \caption{The
experimental setup of a two-arm imaging system based on ghost
imaging. $d_{0}$ is the distance from the light source to an object
as well as to the $\sigma$ plane. A lens with focal length $f_{t}$
and aperture $L_{t}$ is inserted in the test arm (including the
object), and a lens with focal length $f_{r}$ and aperture $L_{r}$
in the reference arm. Both arms are two independent image-forming
systems. $d_{1}$, $d_{2}$, $d_{3}$ and $d_{4}$ satisfy the Gaussian
thin-lens equation: $1/d_{1}+1/d_{2}=1/f_{t}$ and
$1/d_{3}+1/d_{4}=1/f_{r}$.} \label{fig2}
\end{figure}

 On the base of Fig.\ref{fig1}, we add another optical
path and rebuild it into a new two-arm imaging system based on ghost
imaging [see Fig.\ref{fig2}], and here we just consider the case of
thermal light illumination. A beam splitter (BS) behind the thermal
source divides light into two beams propagating through two distinct
arms: in the test arm, a lens with focal length $f_{t}$ is placed
distance $d_{1}$ from an object and $d_{2}$ from a detector $D_{t}$;
in the reference arm, for simplicity assuming a pseudo plane
($\sigma$ plane) at the symmetric position of the object with
respect to BS, a lens with focal length $f_{r}$ is placed distance
$d_{3}$ from the $\sigma$ plane and $d_{4}$ from another detector
$D_{r}$. The relevant distances obey the Gaussian thin-lens
equation: $1/d_{1}+1/d_{2}=1/f_{t}$ and $1/d_{3}+1/d_{4}=1/f_{r}$,
which indicates both arms are two independent image-forming systems,
and are imaging the object and the $\sigma$ plane, respectively.
Although the image of the object can be obtained by the test arm
directly, we pay more attention to the image reconstructed through
the correlation between the two arms. Recording the test arm
intensity $I_{t}(x_{t})$ by $D_{t}$, and correlating it with the
reference arm intensity $I_{r}(x_{r})$ recorded by $D_{r}$, we can
gain information about the object from the correlation function
\cite{Bache04}
\begin{equation}\label{Eq2}
\
G(x_{t},x_{r})=<I_{t}(x_{t})I_{r}(x_{r})>-<I_{t}(x_{t})><I_{r}(x_{r})>.
\end{equation}
In term of results of Ref.\cite{cheng04,Bache04,Gatti06},
Eq.(\ref{Eq2}) can be written as
\begin{equation}\label{Eq3}
\ G(x_{t},x_{r})=\left |\int_{source} dxdx'
  G^{(1)}(x,x')h_{t}(x,x_{t})h_{r}^{*}(x',x_{r})\right|^{2},
\end{equation}
where $G^{(1)}(x,x')$ is the first order correlation function of the
source, and $h_{t},h_{r}$ are the impulse response functions of the
test arm and the reference arm, respectively. Suppose the source is
quasimonochromatic and fully spatially incoherent:
\begin{equation}\label{Eq4}
G^{(1)}(x,x')=I(x)\delta(x-x')
\end{equation}
where $I(x)$ represents the intensity distribution of the source and
$\delta(x)$ is the Dirac delta function.
 Substituting Eq.(\ref{Eq4}) into Eq.(\ref{Eq3}), we have
\begin{equation}\label{Eq5}
 G(x_{t},x_{r})=\left|\int_{source} dx I(x)
h_{t}(x,x_{t})h_{r}^{*}(x,x_{r})\right|^{2}.
\end{equation}

\begin{figure}[htb]
\centering\includegraphics[width=7cm]{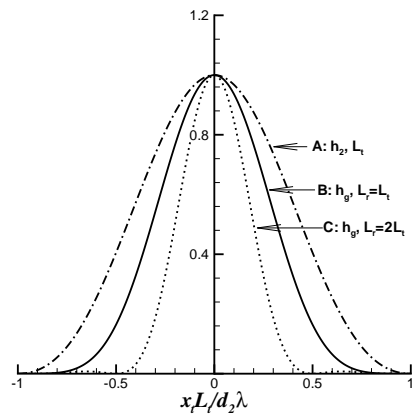} \caption{The
comparison between the FWHMs of $h_{2}(x_{0},x_{t})$ and $h_{g}(x)$
. Dashed line A is the 1-D APSF $h_{2}(x_{0},x_{t})$ of a single
lens with aperture $L_{t}$; solid line B represents the kernel
$h_{g}(x)$ of the two-arm system under $L_{r}=L_{t}$ and Dotted line
C under $L_{r}=2L_{t}$.} \label{fig3}
\end{figure}

  Further, under the paraxial approximation, the impulse response function of the test
arm is given by
\begin{equation}\label{Eq6}
\ h_{t}(x,x_{t})=\int dx_{0}
h_{1}(x,x_{0})t(x_{0})h_{2}(x_{0},x_{t}),
\end{equation}
where $t(x_{0})$ denotes the object transmission function,
\begin{equation}\label{Eq7}
\ h_{1}(x,x_{0})=\frac{e^{jkd_{0}}}{j\lambda
d_{0}}\exp\left\{\frac{i\pi (x-x_{0})^{2}}{\lambda d_{0}}\right\}
\end{equation}
represents free-space propagation from the source to the object, and
\begin{widetext}
\begin{eqnarray}\label{Eq8}
h_{2}(x_{0},x_{t})=\int_
{-\frac{L_{t}}{2}}^{\frac{L_{t}}{2}}dx_{f}\frac{e^{jkd_{1}}}{j\lambda
d_{1}}\exp\left\{\frac{i\pi (x_{0}-x_{f})^{2}}{\lambda
d_{1}}\right\} \exp\left(-\frac{i\pi x_{f}^{2}}{\lambda f}\right )
\frac{e^{jkd_{2}}}{j\lambda d_{2}} \exp \left\{ \frac{i\pi
(x_{t}-x_{f})^{2}}{\lambda d_{2}} \right\} \propto \text{sinc}
\left\{\left(\frac{x_{0}}{d_{1}}+\frac{x_{t}}{d_{2}}\right)\frac{L_{t}}{\lambda}\right\}
\end{eqnarray}
describes the one-dimensional (1-D) amplitude point spread function
(APSF) of the lens of the test arm. $\lambda$ is the source
wavelength, $k=2\pi/\lambda$ is wave number, and $L_{t}$ is the
aperture of the lens in the test arm. Substituting
Eq.(\ref{Eq7})-Eq.(\ref{Eq8}) into Eq.(\ref{Eq6}), we get
\begin{eqnarray}\label{Eq9}
\ h_{t}(x,x_{t})\propto &&\int dx_{0} \frac{e^{jkd_{0}}}{j\lambda
d_{0}}\exp\left\{\frac{i\pi (x-x_{0})^{2}}{\lambda d_{0}}\right\}
t(x_{0})\text{sinc}\left\{\left(\frac{x_{0}}{d_{1}}+\frac{x_{t}}{d_{2}}\right)\frac{L_{t}}{\lambda}\right\}.
\end{eqnarray}
Similarly to $h_{t}(x,x_{t})$, the impulse response function of the
reference arm is directly given by
\begin{equation}\label{Eq10}
\ h_{r}(x,x_{r})\propto \int dx'_{0}\frac{e^{jkd_{0}}}{j\lambda
d_{0}}\exp\left\{\frac{i\pi (x-x'_{0})^{2}}{\lambda
d_{0}}\right\}\text{sinc}\left\{\left(\frac{x'_{0}}{d_{3}}+\frac{x_{r}}{d_{4}}\right)\frac{L_{r}}{\lambda}\right\}.
\end{equation}
where $L_{r}$ is the aperture of the lens in the reference arm. If
the source is infinitely large and the intensity distribution is
uniform, $I(x)=I_{0}$; then substituting
Eq.(\ref{Eq9})-Eq.(\ref{Eq10}) into Eq.(\ref{Eq5})),
 after calculation, we obtain
\begin{eqnarray}\label{Eq11}
\ G(x_{t},x_{r})\propto I_{0}^{2}\left|\int dx_{0}
t(x_{0})\text{sinc}\left\{\left(\frac{x_{0}}{d_{1}}+\frac{x_{t}}{d_{2}}\right)\frac{L_{t}}{\lambda}\right\}
\text{sinc}\left\{\left(\frac{x_{0}}{d_{3}}+\frac{x_{r}}{d_{4}}\right)\frac{L_{r}}{\lambda}\right\}\right|^{2}\nonumber\\
=I_{0}^{2}\left|\int
dx_{0}t(x_{0})\text{sinc}\left\{\left(x_{0}+\frac{x_{t}}{M_{t}}\right)\frac{L_{t}}{\lambda
{d_{1}}}\right\}\text{sinc}\left\{\left(x_{0}+\frac{x_{r}}{M_{r}}\right)\frac{L_{r}}{\lambda
{d_{3}}}\right\}\right|^{2},
\end{eqnarray}
where $M_{t}=d_{2}/d_{1}$ and $M_{r}=d_{4}/d_{3}$ are the
magnifications of the imaging systems in the test arm and the
reference arm, respectively. For a simple case of
$x_{r}=M_{r}x_{t}/M_{t}$, Eq.(\ref{Eq11}) becomes
\begin{eqnarray}\label{Eq12}
G\left(x_{r}=\frac{M_{r}}{M_{t}}x_{t}\right)\propto \left|\int
dx_{0}t(x_{0})\text{sinc}\left\{\left(x_{0}+\frac{x_{t}}{M_{t}}\right)\frac{L_{t}}{\lambda
{d_{1}}}\right\}
\text{sinc}\left\{\left(x_{0}+\frac{x_{t}}{M_{t}}\right)\frac{L_{r}}{\lambda
{d_{3}}}\right\} \right|^{2}
\end{eqnarray}
which represents a special point-to-point intensity correlation and
has the form of a coherent imaging scheme. Its kernel
\begin{eqnarray}\label{Eq13}
h_{g}\left(x_{r}=\frac{M_{r}}{M_{t}}x_{t}\right)=\text{sinc}\left\{\left(x_{0}+\frac{x_{t}}{M_{t}}\right)\frac{L_{t}}{\lambda
{d_{1}}}\right\}\text{sinc}\left\{\left(x_{0}+\frac{x_{t}}{M_{t}}\right)\frac{L_{r}}{\lambda
{d_{3}}}\right\}
\end{eqnarray}
\end{widetext}
is the product of the 1-D APSFs of the two lenses, and analogous to
the APSF of confocal laser scanning microscopy (CLSM) \cite{CLSM}.
As shown in Fig.\ref{fig3}, under the two lenses with same aperture
$L_{r}=L_{t}$, the ratio between the full width at half maximum
(FWHM) of $h_{g}(x)$ (solid line B) and that of $h_{2}(x_{0},x_{t})$
(dashed line A) is nearly 1/1.4, which suggests decreasing the
spatial extent of the diffraction spot and increasing resolution by
a factor of 1.4. What's more, the FWHM of $h_{g}(x) $ can be further
diminished by enlarging the aperture $L_{r}$ of the lens in the
reference arm (dotted line C), which is important to increase
resolution of the two-arm imaging system with a low-NA lens in test
arm.

\begin{figure}[htb]
\centering\includegraphics[width=7cm]{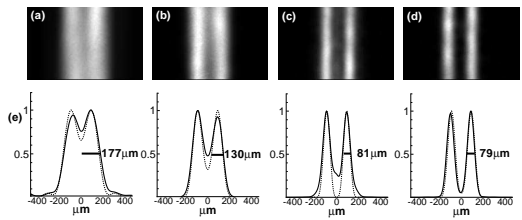} \caption{The
 acquired images of the double slit from the two-arm imaging system. (a) was produced
directly by the test arm under $f_{t}=400 mm,L_{t}=3 mm$, and
(b)-(d) were generated through the correlation between the same test
arm and different reference arms under (b): $f_{r}=400 mm, L_{r}=3
mm$; (c): $f_{r}=400 mm, L_{r}=6 mm$; (d): $f_{r}=250 mm, L_{r}=3
mm$. In (e), solid lines denote the normalized horizontal section of
the images of (a)-(d), and dashed lines are corresponding
theoretical curves.} \label{fig4}
\end{figure}

\begin{figure}[htb]
\centering\includegraphics[width=7cm]{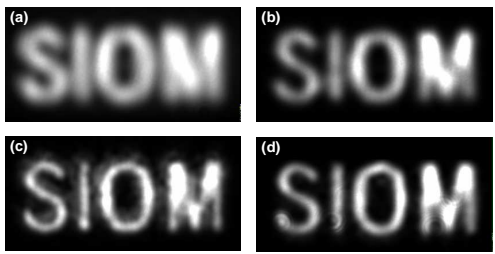} \caption{The
acquired images of the letters \textbf{``SIOM"} from the two-arm
imaging system. The experimental parameters of (a)-(d) are the same
with that of Fig.\ref{fig4}(a)-(d), respectively.} \label{fig5}
\end{figure}

 In the experiment, the thermal source was simulated by the pseudo-thermal light
generated by a frequency-doubled pulsed Nd:Yag laser ($\lambda =
0.532$ $\mu$m) hitting a slowly rotating ground-glass disk, and two
CCD cameras  were used to record the light intensities of both arms,
respectively. We first put a double slit (the slit width 90 $\mu$m
and the center-to-center separation 180 $\mu$m) in the object plane,
and chose two lenses with same focal length ($f_{t}=f_{r}= 400$ mm)
in the two arms. The transmission aperture $L_{t}$ of the lens of
the test arm was fixed at 3 mm  by an iris diaphragm in the whole
experimental process, while the aperture $L_{r}$ in the reference
arm could range from 3 to 20 mm by another iris diaphragm. Taking
$d_{1}=d_{2}=2f_{t}$ and $d_{3}=d_{4}=2f_{r}$ made the magnification
$M_{t}=M_{r}=1$. As a result of Eq.(\ref{Eq1}), the resolution limit
of the test arm is $1.22\lambda d_{1}/L_{t}\approx173$ $\mu$m,
approximately to the double-slit distance 180 $\mu$m. Thus, we only
distinguished the double slit barely by the imaging system of the
test arm and got a blurry image [see Fig.\ref{fig4}(a)]. However,
under the same aperture $L_{r}=L_{t}=3$ mm,  we could gain a
relatively clear image via the correlation between the two arms
[Fig.\ref{fig4}(b)]. Furthermore, a higher-resolution image was
obtained by expanding $L_{r}$ to 6 mm [Fig.\ref{fig4}(c)]. Besides,
keeping $L_{r}=3$ mm and $M_{r}=1$ invariant, we also got a better
image by employing a lens with short focal length ($f_{r}$=250 mm)
in the reference arm [Fig.\ref{fig4}(d)].
The quantitative comparison can be seen from the normalized
horizontal section plotted in Fig.\ref{fig4}(e) (solid line), which
agrees with the theoretical analysis (dashed line). The images of a
more complex object (a mask with letters \textbf{``SIOM"}) were
gained by repeating above experimental processes [see
Fig.\ref{fig5}]. These results show that enhancing the resolving
power of the reference arm where there is no object to be observed,
can increase the resolution of the image effectively.

  It's well known that medical endoscopes are very useful instruments
in disease diagnosis. While the narrow space between human internal
organs only allows the probe with a small lens into the body, which
restricts the image resolution. To overcome the problem, a two-arm
endoscope based on ghost imaging can be developed. Because of no
test objects in the reference arm, the imaging system is not
confined to the endoscopic working environment and may use a larger
lens on the outside of the body to generate higher-resolution images
through the correlation. And for transmission x-ray microscopes, the
transverse resolution is equal to $\beta\lambda/NA_{F}$, where
$NA_{F}$ is the numerical aperture of a Fresnel zone plate and
$\beta$ is an illumination dependent constant \cite{Jaco92}. The
focal depth of the zone plate is calculated by $\Delta
z\approx\pm\frac{1}{2}\lambda/NA_{F}^{2}$, following the definition
of Born and Wolf \cite{Born}. Hence, as one increases $NA_{F}$, the
resolution improves linearly, while the focal depth decreases as the
square that limits the thickness of specimens under investigation.
This dilemma can also be solved by a x-ray microscope with two arms:
in the test arm using a Fresnel zone plate with long focal depth
permits a certain penetration depth, and in the reference arm
selecting another Fresnel zone plate of high NA guarantees required
resolution. The two-arm imaging scheme is also applicable to many
other microscopic systems where the NAs of their objective lenses
are limited.

   In conclusion, we have demonstrated the feasibility of a
microscope scheme based on ghost imaging technique for improving the
resolution of a lens-limited imaging system, and briefly discuss
potential applications of this ``ghost" microscope. Compared with
CLSM, the two-arm microscope system has a similar APSF and also
realizes high resolution, but is more flexible and convenient in
manoeuvring optical components because its test arms and reference
arm are two independent imaging systems.

  This research is partially supported by the Hi-Tech Research and
Development Program of China, Project No. 2006AA12Z115, and Shanghai
Fundamental Research Project, Project No. 06JC14069.

\end{document}